# Laboratory evidence of the halting of magnetic reconnection by a weak guide field

S. Bolanos[1,2], R. Smets[2,*], S.N. Chen[3], A. Grisolet[4], E. Filippov[5,6], J.L Henares[7], V. Nastasa[3,9], S. Pikuz[5,6], R. Riquier[4], M. Safranova[8], A. Severin[1], M. Starodubtsev[8], J. Fuchs[1,+]

[1]LULI - CNRS, CEA, UPMC Univ Paris 06: Sorbonne Université, Ecole Polytechnique, Institut Polytechnique de Paris - F-91128 Palaiseau cedex, France
[2]LPP, Sorbonne Université, CNRS, Ecole Polytechnique, F-91128 Palaiseau, France
[3]ELI-NP, "Horia Hulubei" National Institute for Physics and Nuclear Engineering, 30 Reactorului Street, RO-077125, Bucharest-Magurele, Romania
[4]CEA, DAM, DIF, F-91297 Arpajon, France
[5]National Research Nuclear University MEPhI, 115409 Moscow, Russian Federation
[6]Joint Institute for High Temperatures, RAS, 125412, Moscow, Russian Federation
[7]Centre d'Etudes Nucléaires de Bordeaux Gradignan, Université de Bordeaux, CNRS-IN2P3, Route du Solarium, F-33175 Gradignan, France
[8]Institute of Applied Physics, 46 Ulyanov Street, 603950 Nizhny Novgorod, Russian Federation
[9]National Institute for Laser, Plasma and Radiation Physics, Magurele, Ilfov, Romania

*: roch.smets@upmc.fr
+: julien.fuchs@polytechnique.fr

ABSTRACT

Magnetic reconnection occurs when two plasmas having co-planar but anti-parallel magnetic fields meet. At the contact point, the field is locally annihilated and the magnetic energy can be released into the surrounding plasma. Theory and numerical modelling still face many challenges in handling this complex process, the predictability of which remains elusive. Here we test, through a laboratory experiment conducted in a controlled geometry, the effect of changing the field topology from two-dimensional to three-dimensional. This is done by imposing an out-of-plane (guide) magnetic field of adjustable strength. A strong slowing down or even halting of symmetric reconnection is observed, even for a weak guide-field. Concomitantly, we observe a delayed heating of the plasma in the reconnection region and modified particle acceleration, with super-Alfvenic outflows ejected along the reconnection layer. These observations highlight the importance of taking into account three-dimensional effects in the many reconnection events taking place in natural and laboratory environments.

I. Introduction

Magnetic reconnection is the subject of intense investigations due to its suspected role in the dynamics of many spatial and astrophysical events, e.g. solar flares[1], planetary magnetic substorms[2], or plasma jets[3]. There, this phenomenon is frequently invoked to explain sudden plasma heating or particle energization, even though the abrupt pace at which these unfold remains difficult to explain. Early on, in the frame of resistive magnetohydrodynamics (MHD),

Sweet[4] and Parker[5] have been able to predict a reconnection rate for two plasmas having their magnetic field lying in a purely two-dimensional plane. Their generalized model has been shown to adequately explain reconnection in collisional plasmas[6]. More recently, the integration of the resistive tearing instability, as an additional effect into the picture, was demonstrated to improve prediction of fast reconnection even for low collisionality plasmas[7,8]. However, despite this and further theoretical effort from many groups (see e.g. Refs[7–10]), a persistent difficulty lies in the inadequacy of models in being able to predict accurately the onset and temporal evolution of magnetic reconnection in most conditions. Indeed, whatever the collisionality of the plasma is, much faster reconnection is still observed in various events compared to that modelled. This has also motivated the development of laboratory experiments dedicated to investigate and understand this process, using e.g. magnetic[11,12] or inertial[13] confinement, or pulsed-power driven plasmas[14].

Aside from elucidating the source of this persistent difficulty, e.g. the exact role played by electrons[9] or ions[10] in the microphysics of reconnection, another factor complicating the picture is the topology of the fields. Deviating from the idealized two-dimensional picture of a canonical reconnection event, the magnetic fields are frequently dynamically evolving in three dimensions. In this case, the presence of a so-called guide field, i.e. normal to the plane of the plasmas and initial magnetic fields, has been evoked for some time to possibly impact the structure of reconnection region and its governing microphysics[15,16]. Examples near us are on the Earth's dayside magnetopause[17] and in the solar corona[18]. Such guide field could not only affect the rate at which reconnection unfolds, but also the directionality of the particles accelerated following reconnection. However, its exact role and influence are still debated. For strong guide field, i.e. when its strength is comparable or larger than the one of the in-plane magnetic fields, opposite results have been highlighted, from quenching reconnection[19–21] to conversely aiding fast reconnection to take place[18]. For weaker guide field, the reconnection rate has not been evoked to be affected[16,22], or only weakly so.

Here we demonstrate in the laboratory, and in a controlled geometry, that imposing a weak guide field on an in-plane, symmetric, reconnection topology results in a strong slowing down, even halting, of the reconnection process. This is also shown to strongly impact the directionality of the particles that are energized in the process.

As shown in Fig.1, this is done by using two high-power lasers beams irradiating solid targets[13,23–27], which creates two hot, dense (see Supplementary notes 1-3) adjacent plasmas, expanding toward each other supersonically (at 81 km/s)[28]. Each is surrounded by a magnetic ribbon[29], i.e. a rather flat toroid magnetic field that is compressed against the target[28,30], which has a ~300 T strength. We note that such setup has limitations in the lifetime (ns-scale) and spatial extend (mm-scale) of the magnetic ribbons that can be created compared to larger-scale setups[31,32], but that it is free from any surrounding structures, such as coils, that can affect the dynamics of the plasmas in pulse-power driven machines[32].

As a result of the convergence of the two plasmas (in less than 1 ns, see Supplementary note 4), a near-ideal bi-dimensional reconnection event can take place, the relevance of which to solar events has been suggested[33,34]. Now, by tilting one target with respect to the other (see Fig.1), an out-of-plane guide field naturally arises over the region where the plasmas encounter.

Our method for diagnosing the evolution of the reconnecting fields uses fast, laminar protons (see Methods) that allow to obtain 2-D snapshots of the magnetic fields over time. These images are then compared to synthetic images generated by using numerically simulated magnetic fields. The unambiguous features observed in both experimental and synthetic images allow us to identify the various phases of reconnection as it unfolds.

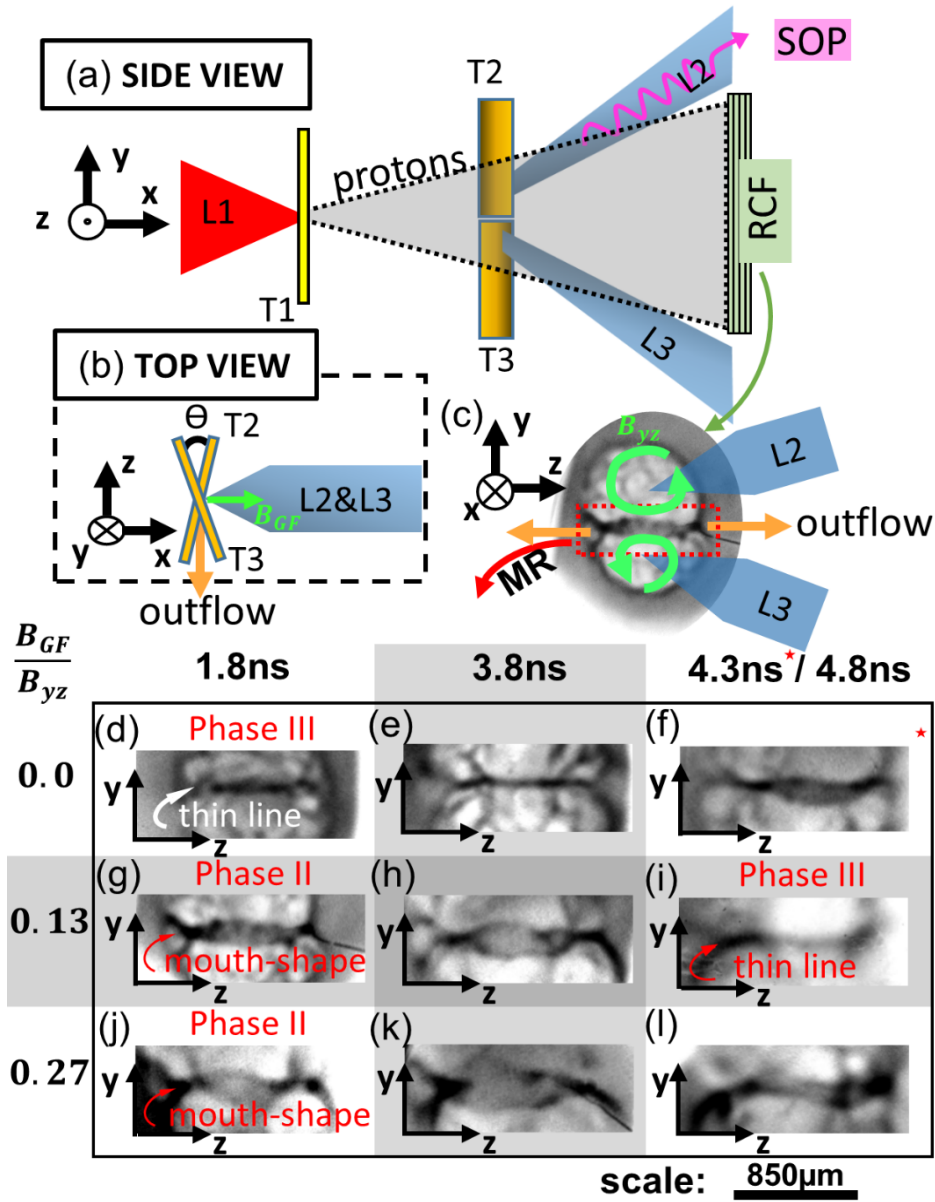

Figure 1: Experimental setup and experimental observation of delayed magnetic reconnection in the presence of a guide field. (a-b) Two schematic views of the experimental setup, along two projections. The high-intensity laser beam L1 (red cone) generates a proton beam (grey cone) from a 25 µm thick Au foil (T1, in yellow). After a 1 mm gap, this proton beam propagates through the two 5 µm thick Cu interaction foils (T2 and T3, coloured in orange). These are irradiated by two high-energy laser beams L2 and L3 (blue cones), which are separated by 500 µm. At a distance of 9 cm after crossing targets T2 and T3, the protons are collected on a film (RCF, green) stack. A variable angle $\vartheta$ can be set between T2 and T3, as shown in panel b, allowing to induce a guide field ($B_{GF}$), aligned with the bisector of the normals to T2 and T3. Note that this is done by tilting each target by $\vartheta/2$, so that the tilt is distributed symmetrically between the two targets, and such the guide field, oriented along x, does not directly affect the deflection of the probing protons. The strength of this guide field is controlled by modifying $\vartheta$ (see Methods). The purple wavy arrow indicates the direction in which the optical self-

*emission is recorded, while the orange arrow indicates the direction of the particle spectrometer in which the ejected particles (ions and electrons) are analyzed (see Methods). (c) Typical image of an RCF film taken at 1.8 ns; the magnetic fields present on the surfaces of T2 and T3 have induced deflections upon the propagating 14 MeV protons and hence proton dose modulations on the film. The dark regions correspond to an accumulation of protons while the light regions correspond to a depletion of protons. With the probing geometry shown in (a-b), the observed outward expulsion of protons corresponds to a clockwise magnetic field ribbon (as indicated by the green arrows) induced on each target by the lasers. Note that the ribbon-like magnetic fields exist only as long as the laser irradiation is maintained; afterward, they quickly disassemble with the plasma. The red dashed rectangle points where the two magnetic ribbons are compressed against each other and where magnetic reconnection (MR) occurs. (d-l) Zooms in the MR region of the proton deflectometry images, at various times and for various strengths of the guide field ($B_{GF}/B_{yz}$ gives the ratio of the guide field vs. the in-plane magnetic fields), as shown. The case $B_{GF}/B_{yz}$ =0 corresponds to the case where the two targets are coplanar. For all images, time 0 corresponds to the start of the targets irradiation by L2 and L3. The spatial scale at the bottom applies for all images and is relative to the target plane.*

The temporal and spatial dose modulations imparted on the probing protons, such as the ones shown in Fig.1.d-l which display magnified views in the zone of contact between the two magnetic ribbons, allow us to analyse the temporal dynamics of how the overall magnetic field topology evolves.

A clear feature of the coplanar case (Fig.1.d-f) is the quick appearance (in less than 1 ns, see Fig.1.d and Supplementary note 4) of a thin line of a compressed probing protons in the plasma encounter area. As will be detailed later in comparison with the synthetic images generated from the numerically simulation, this line reveals the thinning of the current sheet associated with the onset of reconnection and of field annihilation. Hence, we witness that reconnection takes place quickly in the coplanar configuration. The appearance of this thin line is however significantly delayed at late times (see Fig.1.i) when applying a weak (0.13) guide field. Rather, at early times, the probing protons form much wider, "mouth"-shaped pattern, that testified of an enhanced deflection compared to that of the coplanar case. This is due to magnetic field pile-up in the area as the magnetic field cannot get annihilated through reconnection and as magnetic field flux is constantly coming into the area (see Supplementary note 1), fed by the constant laser energy deposition at the two irradiation spots. As the guide field strength is increased, the "mouth" widens, as reconnection becomes more difficult and magnetic field accumulation increases; for $B_{GF}/B_{yz}$ =0.41, we do not even witness the onset of reconnection during the magnetic field lifetime (see Supplementary note 4).

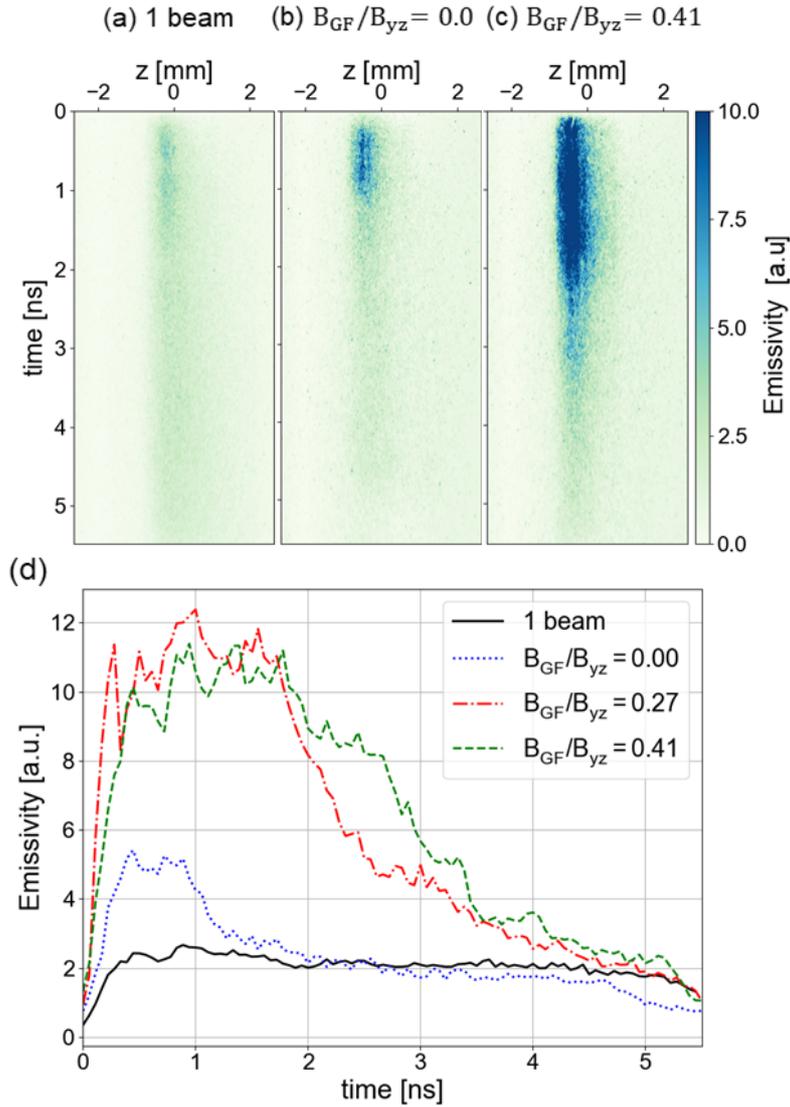

*Figure 2: Laboratory optical pyrometry observation of delayed heating in the reconnection area and in the presence of a guide field. (a-c) streaked images of the self-emission of the plasma in the reconnection region as recorded in the laboratory experiment along the z-axis (see Fig.1), in the mid-plane between the irradiation spots of the L2 and L3 lasers. The self-emission is recorded for photons around (470±135) nm wavelength and integrated over 230 μm along the y-axis (see Fig.1). Panel a corresponds to only one laser beam (L2) turned on. Panels b and c correspond to the case where L2 and L3 are fired; for panel b on coplanar targets, while for panel c one of the target is tilted by 45°, resulting in the presence of a $B_{GF}/B_{yz}$ =0.41 guide field. For all, we observe that at late times, past 5 ns, the signal dies out. Indeed, as the laser is switched off, the heat flux cannot maintain the magnetic field anymore. As a consequence, the dense plasma inducing the observed emission cannot be confined anymore in the magnetic field and quickly expands into vacuum. (d) Lineouts of z-integrated streaked plasma self-emission and as a function of time of the images shown in (a-c), plus of another shot corresponding to an intermediate guide field strength. Time t=0 corresponds to the start of the targets irradiation by L2 and L3.*

The time-resolved self-emission observed to originate from the reconnection region (see

Methods), shown in Fig.2, concurs with the observation of delayed reconnection inferred from the proton deflectometry images shown in Fig.1. The emitting plasma is in an optically thin regime (see Methods), meaning that the self-emission increases with the plasma density, but reduces with the plasma temperature. When only one laser beam (either L2 or L3) irradiates the target assembly, we observe in the reconnection region, i.e. 250 µm away from the laser spot, a quite steady self-emission over time (see Fig.2.a and the full line in Fig.2.d). This takes place with a slight delay (~0.3 ns) compared to the start of the laser irradiation, this delay being due to the need for the plasma to expand laterally up to the location of observation.

The overall behaviour is quite different when the two laser beams irradiate the targets. At the onset of the self-emission, we observe a first increase of the self-emission compared to that induced by one laser beam. It is followed by a plateau, and, later, a decrease (see Fig.2.d). We interpret the first, fast increase as due to the increased density in the reconnection region induced by the pile-up when the two expanding plasma collide there. The later decrease of the self-emission in the two beams case is likely due to two cumulated factors: (i) as reconnection takes place, the accumulated plasma can be evacuated from the reconnection layer, and (ii) the increased temperature of the plasma as the magnetic energy is transferred to the plasma. Without guide field, this emission decrease is seen in Fig.2.d to take place around 1 ns, which is consistent with the onset of reconnection in this case (see Supplementary Fig. 8). In the presence of a guide-field, we notice three changes: the first increase is enhanced, the plateau lasts longer and the decrease takes place later, the delay in the latter increasing with the guide-field strength (compare the dashed and dotted-dashed lines in Fig.2.d). All this is well-consistent with the delayed reconnection of the magnetic field observed in Fig.1: both the evacuation of the piled-up plasma and its heating are delayed.

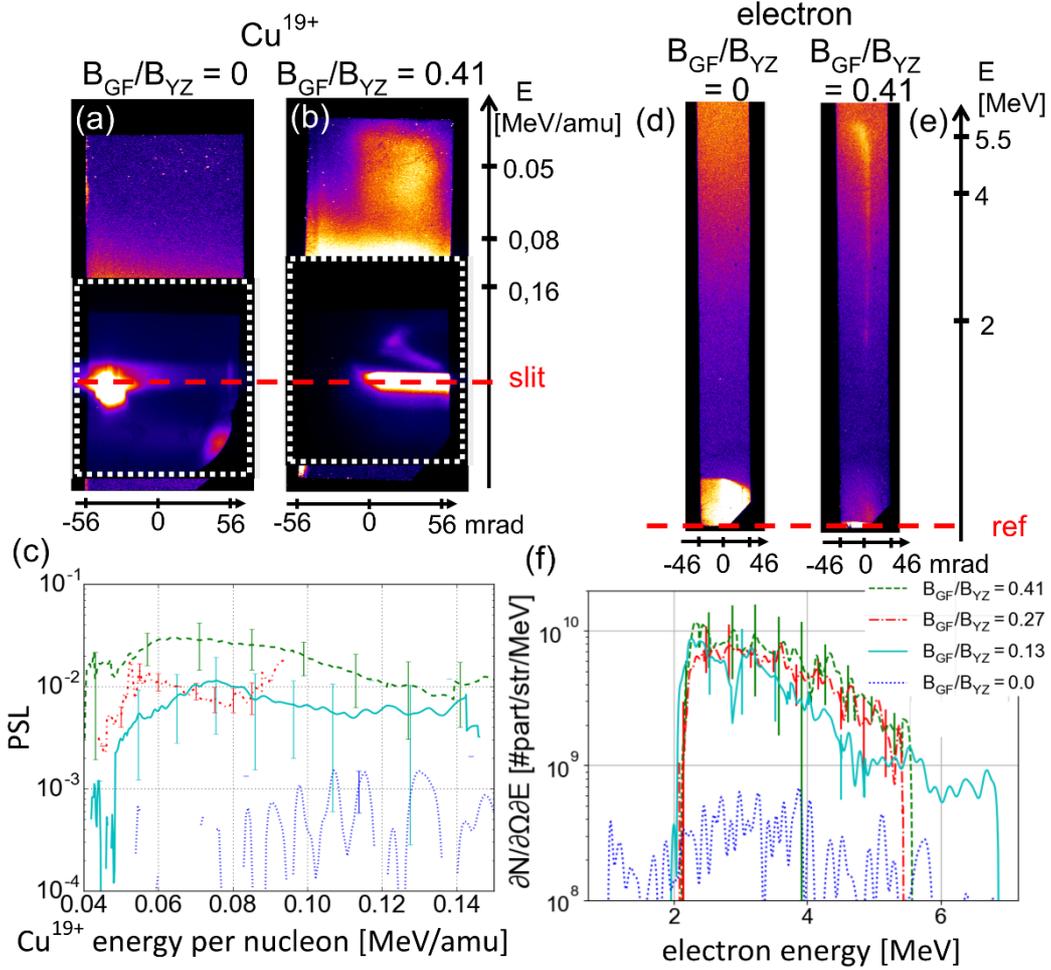

*Figure 3: Laboratory evidence for enhanced particle acceleration along the current sheet in the presence of a guide field. (a,b) Raw spectra of Cu ions as recorded by the particle spectrometer in the absence (panel a) or presence (panel b) of a guide field. Since our targets are made of Copper, note that we made the assumption that the ions recorded in the spectra are $Cu^{19+}$, based on the average ionization state we expect to have in our plasma conditions (see Supplementary note 6). The vertical axis is the spectral one, the horizontal one is sampling the angle of emission of the particles from the targets. For panel b, we have $B_{GF}/B_{yz} = 0.41$. In the region surrounding the projection of the spectrometer entrance slit onto the detector, the images are strongly saturated by the X-ray and visible light emitted from the plasma. This is why the images shown are a patch of the first scan of the image plate detector (top part) where the ion signal can be seen with the second, unsaturated and delimited by the white dashed line, scan of the detector. Because of the noise close to the slit, the spectrometer was reliably able to detect ions only in the shown energy range of 0.04 keV/amu to 0.16 keV/amu. (c) Lineouts of the spectra such as shown in (a), without guide-field (dotted line) or with a guide-field of various magnitude (full, dashed and dotted-dashed lines, see panel f for the legend). Each plot results from the integration over the angular dimension and is obtained by averaging two to five shots, depending on the configurations, recorded in the same conditions. The error bars corresponding to the standard deviation of the signal over these shots. (d,e,f) Same as (a,b,c) for the electrons. We note that the recorded energies are much higher than if the particles would be ejected merely at the Alfven velocity, as predicted in the Sweet-Parker resistive MHD model of reconnection. However, this is not surprising as already many*

*numerical and experimental studies have shown that higher energy gain could be expected*[35–38].

Figure 3 shows that when a guide field is applied to our reconnection setup, strong ion and electron spectra are recorded (see Methods) being ejected along the current sheet axis (z-axis), i.e. along the expected outflow direction (as indicated in Fig.1). Quite differently, no signal in both channels can be recorded about the noise level in the absence of guide field, or when looking in the perpendicular direction (see Supplementary note 6).

The absence of signal in our case in the coplanar case is not so surprising: since our spectrometer looks in the reconnection plane, along the target surface, it would miss the particles that are accelerated out-of-plane. This is likely the case for most of the particles accelerated *following reconnection*: as the particles in the plasma inflows reach the reconnection area, they will be influenced by the $E_x$ component of the electric field. Since this component is normal to the reconnection plane, it will skew their trajectories out-of-plane[39,40]. The interesting point is that a strong signal is seen in the presence of a guide field, i.e. when the reconnection is slowed down and that magnetic field is compressed and accumulated on both side of the reconnection layer. These particles are thus likely not accelerated during reconnection, but ahead of the actual reconnection, either through slingshot Fermi acceleration[39] or betatron acceleration[40,41]. Since this requires an accumulation of magnetic field powering the acceleration, the observation of energization when the guide-field is applied is well compatible with the observation of piling-up of magnetic field in that configuration compared to the coplanar one.

These findings are now compared to numerical simulations performed using the HECKLE hybrid code (see Methods) in the conditions of the experiment. The results of two simulations are shown. Fig.4 shows the result of a simulation conducted using a ratio between the plasma and magnetic pressure (β= $p_{plasma}/p_{magnetic}$ ) =1, while Fig.5 corresponds to β =20. For β =1, reconnection unfolds at a slow pace and thus allow us to clearly identify the different phases of the reconnection process. Panel (i) of Fig. 4 displays the time evolution of the reconnection rate (black line) and the associated reconnected flux (red line) where we can identify four phases: (I) initialisation, (II) onset of the reconnection, (III) during the reconnection and (IV) after the reconnection. To make clear these phases, panel a-d of Fig. 4 portray the density profile of the overall plasma in greyscale, and the associated in-plane magnetic field lines. At the same time, panel e-h provide the proton dose simulated using the ILZ code (see Methods). Using the simulated electric and magnetic field provided by the simulation and allowing us to interpret the time evolution of the proton dose displayed in Fig. 1d-l. In phase I, the two magnetic ribbons are not yet in contact, and the resulting proton deflectometry image is just the linear overlap of the two images that are produced by the two independent ribbons. In phase II, the magnetic fields start to compress against each other, resulting in a thinning of the current sheet, just before the onset of reconnection. The increased magnetic field at the contact point leads to an increased deflection of the probing protons, resulting in a central white zone that has the characteristic shape of a "mouth". In phase III, the magnetic fields start to annihilate, resulting in a lesser deflection of the protons, and hence to the central characteristic black thin line in the proton images. In phase IV, the anti-parallel magnetic fields have been fully annihilated in the central zone, the magnetic field have been rearranged as to just surround the overall structure and no significant proton deflection takes place anymore in the central region.

Comparing Fig.1 and 4, we can readily observe that indeed in the coplanar case (Fig.1.d-e), the reconnection takes place quickly. After that, the magnetic field topology in the interaction region is not observed to significantly change over time since the width and dose (with respect to the background (unperturbed) proton beam) of the thin line of focused protons appears stable (see Supplementary note 5). This suggests that the magnetic field, which is permanently generated by the Biermann-Battery mechanism, is simultaneously annihilated through reconnection, consistently with what was previously observed[42] in a similar coplanar geometry and with similar laser parameters. Using the characteristics of the focused proton line, we can estimate the magnetic field in the reconnection region, either analytically or through simulating the proton deflectometry maps, as in Fig.4.e-h (see Supplementary note 5). This yields, in each ribbon, an integrated (along the axis x) strength of the magnetic field ~3 T.mm, and a width (along the axis y) ~150 µm. Finally, at late times (4.3 ns, see Fig.1.f), the central proton line widens, meaning that the magnetic field compression increases, which is likely due to a slowing down of the reconnection rate toward the end of the magnetic field lifetime.

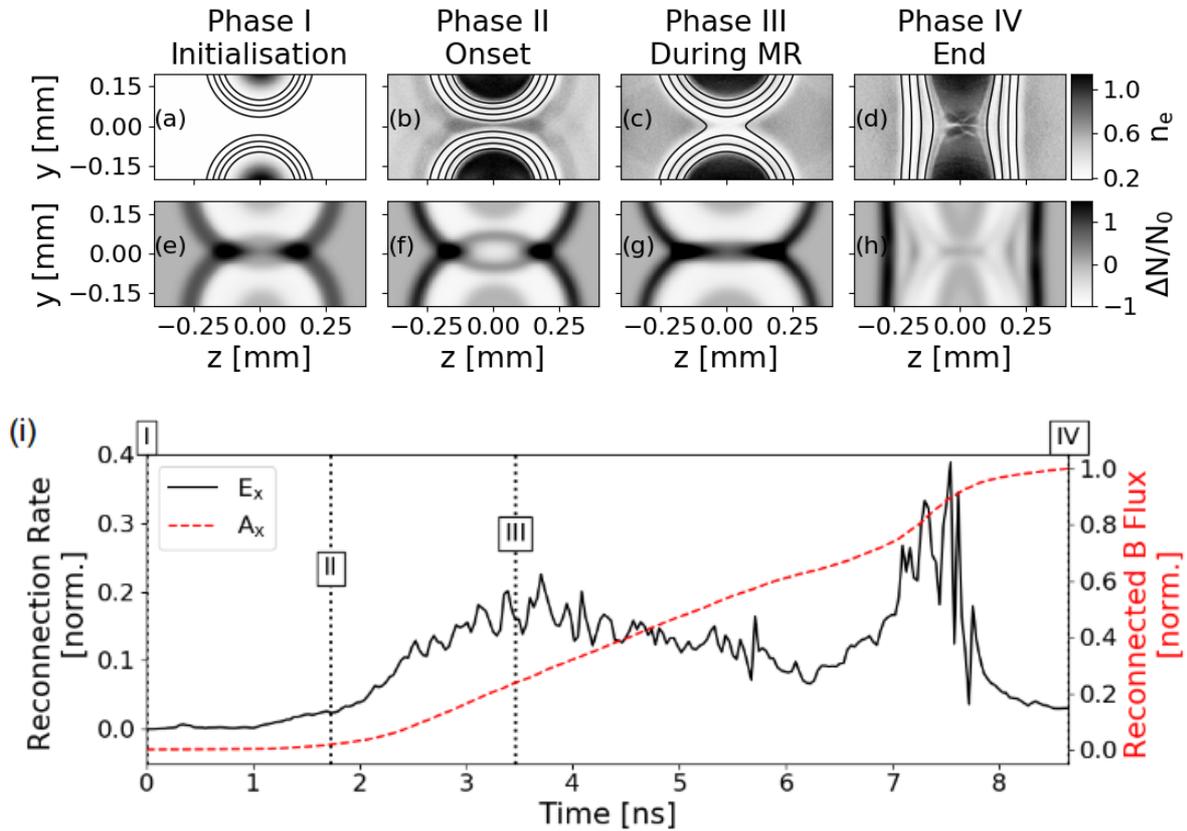

*Figure 4: Hybrid simulation of a $\beta$ =1 magnetic reconnection event allowing for the identification of its different phases in the proton deflectometry images. (a-d) Snapshots extracted from a two-dimensional hybrid simulation (see Methods) of the reconnection between two plasmas expanding toward each other and in which magnetic toroids are embedded; the lines represent the isocontours of the magnetic field, while the grayscale represents the normalized electron plasma density ($n_e$). The different phases of the reconnection are identified above the panels. (e-h) Synthetic images (see Methods) of proton-deflectometry using the magnetic field distribution shown in (a-d). These are computed for 14 MeV protons, consistently with the experimental images shown in Fig.1. The greyscale*

*represents the relative proton dose modulation ($\Delta N/N_0$) induced by the protons passing through the field structures shown in (a-d). (i) Temporal evolution of both the reconnection rate (calculated here as the out-of-plane $E_x$, normalized by the Alfven velocity and the maximum magnetic field[9], full black line) and of the reconnected magnetic flux (red dashed line). As the maximum magnetic field decreases with time, this can induce large spikes in the reconnection rate at late times, as can be seen after 6 ns. Time t=0 corresponds to the start of the simulation, which coincides with the plasma generation and expansion.*

When applying a guide field, the experimentally observed proton pattern is clearly alike that of phase II: at the same times when the reconnection was fully engaged in the coplanar case, it is just at its onset, with the archetypal "mouth" shape up to quite late (3.8 ns, see Fig.1.h and k). A detailed analysis (see Supplementary note 5) shows how much magnetic field piled-up and compressed in the guide field case: the x-integrated strength of the magnetic field grows for example from ~3 T.mm to >12 T.mm in 4 ns for $B_{GF}/B_{yz}$= 0.13.

A similar temporal evolution, but at a much quicker pace, is observed in the simulation performed at β =20. In these conditions, as portrayed in panel a of Fig. 5, the phase during which the current sheet is build-up and thinned (the analogue to phase II in Fig.4) is longer than in the β =1 case. As a consequence, the upstream magnetic field is increased by compression and flux conservation (the magnetic field being frozen-in far enough from the electron diffusion region). Furthermore, the proton Larmor radius increasing with β, a larger amount of them will escape from the ribbon by finite Larmor radius effect, hence decreasing the proton density embedded within the magnetic ribbon. These two facts result in an increased Alfven velocity for the inflow in the reconnection layer, and consequently in a strongly reduced duration over which reconnection takes place. This numerical picture is very different from what happens in the experiment. This can be understood as the simulation contains a finite amount of magnetic flux given as initial condition, while the magnetic flux embedded in the laboratory ribbon is continuously fuelled during reconnection by the laser energy deposition on target. Fig. 5c shows the simulation results obtained with a guide field $B_{GF}/B_{yz}$ = 0.41. Note that the reconnection process is similarly impulsive with and without a guide field, and at a very comparable rate. But it also clearly appears, that the thinning phase with a guide field is significantly delayed compared to the case without a guide field. We ascribe the delay effect to the distortion of the so-called Hall (out-of-plane) quadrupolar magnetic field, as evidenced by comparing the maps of this component shown at the same time in Fig. 5b and d. This field is generally believed to be induced by the growth over time of an in-plane Hall electric field, which is induced by the out-of-plane current resulting from the pinching toward each other of the anti-parallel magnetic fields. Hence, the quadrupolar magnetic field is commonly thought to be a consequence of a reconnection event[44]. However, we have recently suggested[45] that the reconnection process and the appearance of the quadrupolar magnetic field are rather both consequences of the formation of a thin (of the order of the proton inertial length) non-flat current sheet in between the two compressed, anti-parallel magnetic fields. In fact, in high-energy-density conditions such as the ones investigated here, the quadrupolar magnetic field appears to be even a necessary precursor to the event related to the formation of the current sheet. Hence, by imposing a guide-field, the growth of the quadrupolar magnetic field is destabilized and, as a consequence, reconnection is delayed.

These numerical results are in total agreement with previous studies using spacecraft observations at the dayside magnetopause under different β conditions; the reason being still unclear, observations show that at high β values, current sheets are seldomly reconnected[46,47] compared to the β =1 case. Further numerical investigations are needed to decipher the role of the current sheet thickness, the associated topology of the magnetic field and the role of the larger Larmor radius.

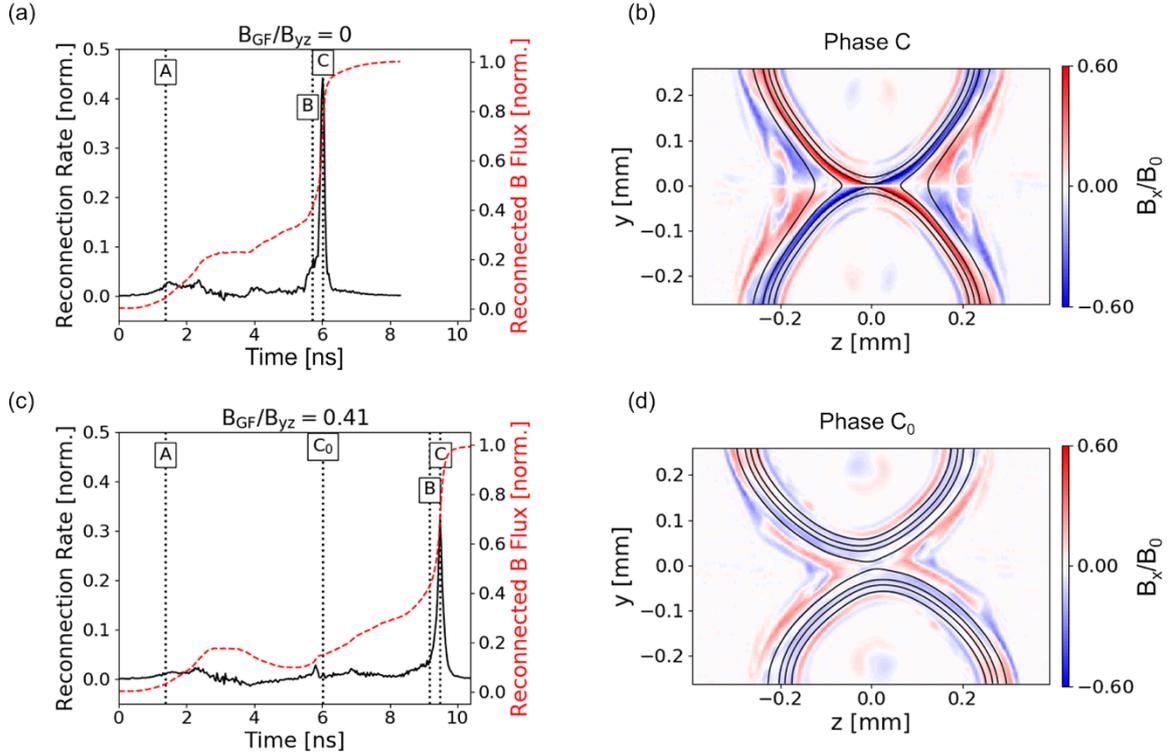

*Figure 5: Simulations of β =20 magnetic reconnection events concurring with the experimental observation of delayed reconnection in the presence of a guide-field. (a) temporal evolution, extracted from the two-dimensional hybrid simulation, of both the reconnection rate (full black line) and the reconnected magnetic flux (red dashed line), together with the identification of the phases of the reconnection (following Fig.4), in the coplanar case ($B_{GF}/B_{yz}$ = 0). (b) snapshot taken during the phase C (at 6.0 ns) of the two-dimensional plane of the simulation; the lines represent the isocontours of the magnetic field, while the colour scale represents the quadrupolar (out-of-plane) component of the magnetic field, still in the coplanar case. (c-d) same as panels a and b, respectively, for the simulation performed in the presence of a guide field ($B_{GF}/B_{yz}$ = 0.41). The phase identified by "$C_0$" in panel c identifies the time of reconnection in the coplanar case. It clearly shows that the onset of reconnection is strongly delayed, and that the quadrupolar magnetic field is much weaker, due to the presence of the guide field. In panel d, one can note the anti-clockwise rotation of the ribbons. This azimuthal drift of the plasma results from the diamagnetic drift of the plasma given by the cross product of the radial pressure gradient (mainly because of the density gradient) by the axial magnetic field (associated to the guide field).*

By means of laboratory experiment performed in a controlled, simplified geometry and free from perturbations induced by the setup, we put forward in this paper how magnetic reconnection can be delayed, or even halted, in the presence of a guide field. We note that these experimental conditions are not only very close to the ones at the Earth magnetopause[47], where satellite observations also reveal that current sheet are less prone to reconnection at large β values, but also to the ones found in reconnecting solar arches[48]. In the solar corona, where reconnection occurs continuously, the structure of the arches is very three-dimensional and contains many sub-substructures. Their encounter is however very difficult to observe because of the involved small scales, complex geometry and overlapping features. Thus, the influence of a guide field, and more generally of a changing three-dimensional topology, is difficult to assert solely from observations, and no conclusive answers could be drawn on the subject. In this frame, by revealing the importance of even a weak guide field, our results could allow to improve the understanding of these reconnection events by adding constrains to their analysis. A useful future direction to further improve our understanding will be to perform similar experiment, but involving several reconnection sites at once, and having them asymmetric[42], as this is also a prevalent condition found in natural events.

**Methods**

Laser experiment
The experiment was performed using the LULI2000 laser facility at the LULI laboratory (France). Three laser beams are used: beams L2 and L3 (see Fig.1) are 5-ns long (square temporal shape). These two beams irradiate simultaneously (within 100 ps) two Copper foils of 5-μm thickness. As shown in Fig.1.b, the two foils were tilted, one by θ/2, the other by -θ/2, such that overall the azimuthal magnetic field ribbon created on one target was tilted by an angle θ with respect to the the other magnetic ribbon created on the other target. Such configuration allows us to have, along the x-axis, a component of the magnetic field that is out of the reconnection plane (the yz plane), leading to the generation of a guide field. We use this setup to generate the guide field rather than an externally imposed magnetic field[49] as the latter would limit us to very weak (tens of T) fields. We studied the influence of the guide field in 4 different cases: co-planar (θ=0°), 15°, 30° and 45° which correspond respectively to the following ratios $B_{GF}/B_{yz}$ of the guide field strength ($B_{GF}$) over the one of the magnetic field in the reconnection plane ($B_{yz}$): 0, 0.13, 0.27, and 0.41.
The two laser focal spot foci are separated by 500 μm for the data shown in Fig.1. Each laser beam has at 1.064 μm wavelength and 200 J of energy. They are equipped with random phase plates in front of the focusing lenses, such that the focal spot is 80 μm in diameter with a uniform intensity distribution. As a result, the intensity on the solid targets was I ~2.5x10$^{14}$ W/cm$^2$.

Diagnostics
The main diagnostic is proton deflectometry. It has been implemented in order to observe the expansion of the two magnetic ribbons induced by L2 and L3 irradiating targets T2 and T3, and the change in the overall magnetic field topology after the two magnetic ribbons interact. Since the magnetic fields produced by the L2 and L3 lasers are mostly contained on the surface of the targets T2 and T3 (see Supplementary note 1), the probing protons are sent quasi-parallel to the normal of the targets (see Fig.1.a-b), such that the deflection are induced by the Lorentz force associated with the probed magnetic fields can be recorded[13,50]. The probe proton dose profile, initially quite uniform[51] after they have left the source target, is modulated due to these deflections and its projection is recorded on films. As the magnetic fields are oriented anti-clockwise with respect to the normal of the target facing the irradiating laser pulse (L2 or L3)[13,28,30], the probing protons are sent through the back of targets T2 and T3 such that the Lorentz force deflection imparted on them is outward. This leads to the observed evacuation of protons (white area) from the centre of each laser irradiation spot, and to a dark rim at the edge, synonym of proton accumulation there, as can be seen in Fig.1.c. To produce the probing proton beamlet, we took advantage of the picosecond-duration laser beam of the LULI2000 facility (L1 in Fig.1). The facility provides such laser beam with an energy on target close to 60 J, having a diameter spot of 10 μm. Hence, its intensity on target is I~ $10^{19}$ W/cm$^2$, allowing us to generate a proton source stemming from the rear surface of the foil as accelerated by the TNSA mechanism[52]. Such a mechanism generates a proton beamlet which energy spectrum is characterised by a 100% dispersion and a maximum energy cut-off of the order of 20-30 MeV, as varying from shot-to-shot. Due to the small source size of such TNSA-accelerated proton beam (of the order of a few microns[53], this method allows us to probe the magnetic fields with an excellent spatial resolution, here limited not by the proton source size, but by the multiple Coulomb scattering they are subject to when crossing the solid targets (see Methods). In practice, the spatial resolution in the plane of targets T2 and T3, is 120 μm. Another advantage of the TNSA-accelerated proton beam is that the maximum aperture of the proton beam can reach a half-angle of 20°[54], which allows to have a compact system: the source of protons can be close of the object to be probed while ensuring to cover a wide field of vision. For example, the magnetic ribbons generated by the Biermann-Battery effect shown in Fig.1 are of the order of 1 to 2 mm in diameter. To radiograph them, it was enough to place the proton source at r=1 cm from the magnetized plasma and the detector was positioned on the other side at l=9 cm. The magnification of the projection onto the detector is thus M = (l + r)/ r = 10.

Due to the short duration of the proton acceleration process (of the order of a picosecond[55]), the temporal resolution is mostly limited by the range of proton energies that are recorded in each of the detector films, that range varying across the film stack. In practice, the films shown in Fig.1 correspond to 14±0.5 MeV, yielding a temporal resolution of 6.9 ps, which is negligible compared to the nanosecond time-scale of the fields evolution. To investigate the temporal evolution of the topology of the magnetic field induced by the high-power laser beams (L2/L3), we simply delay over various shots the ps-laser beam with respect to the ns-duration laser beams in controlled steps, as shown in Fig.1. We verified that the morphology of the proton deflectometry images, at a given proton probing time, was well reproducible from shot to shot in order to reconstruct the temporal evolution of the fields over different shots. After propagation through the solid targets, the protons are recorded on a stack of radiochromic films[56] placed 9 cm away from the solid targets. The stack is composed of HD-v2 and EBT3 films, with a 12 μm-thick Al foil upfront to protect the films from light and debris.

To analyze the x-ray emission produced by the plasmas (detailed in Supplementary note 3), and infer, with support from the hydrodynamic simulation FCI2 (see Supplementary note 1), their temperature, a pair of focusing spectrometers (FSSR)[57] was used. Each spectrometer used a spherically bent mica crystal with parameters *2d = 19.9376 Å* and radius of curvature R = 150 mm. They were implemented to measure the x-ray spectra of multi-charged Copper ions in the range of 9.0–9.5 Å (1300–1380 eV), in the second order of reflection, with a spatial resolution of about 35 μm. The spectrometers were multiplexed such that the spectrometer S1 (resp. S2) provided spatial resolution in the equatorial (resp. vertical) plane. The x-ray spectra were recorded on passive detectors, namely TR Fujifilm Image Plate, protected from optical radiation by thin Polypropylene filters (1 μm thick) covered a coating of 200 nm thick Al. All data were time-integrated.

To diagnose the self-emission of the plasma with temporal resolution, we implemented an optical system to image the plasma, along the axis z, on the entrance slit of a streak camera (model S20 from Hamamatsu). As pictured in Fig.1, the imaging of the plasma is performed through the lens focusing the laser beam L2 on target and in reflection of a thin pellicle that is positioned upstream of that lens. A colour glass filter (BG38 from Schott) is used to filter out the short wavelengths of the emitted spectrum; the long wavelength part of the spectrum is equally removed from the measurement due to the strong decrease of sensitivity of the camera in that region. As a result, we record light in a spectral range of (470±135) nm. The output of the diagnostic is an image which provides, on one axis, spatial resolution along the current sheet as well as, on the other axis, temporal resolution, as shown on the raw data of Fig.2.a-c. Based on hydro-radiative simulations of our experimental configuration (detailed in Supplementary note 1), we analyse that the maximum of emissivity originates from an optically thin plasma: in the reconnection zone, where the electron density is of the order of $10^{20}$ cm$^{-3}$, the mean free path of a visible photon is 46.1 µm, which is of the same order than the plasma gradient scale length 20-40 µm. As a consequence, Bremsstrahlung is the dominant emission mechanism in the spectral observation range of the diagnostic[58]. Hence, the emissivity of the plasma can be expressed as: $E_{ff} \sim Z \times n_e^2/[(T_e)^{1/2} \times g_{ff} \times \nu]$, where $g_{ff}$ is the velocity averaged gaunt factor[59], and $\nu$ is the spectral bandwidth of the diagnostic.

The particle spectrometry (see also Supplementary note 6) is performed simultaneously on ions and electrons by using different detectors recording positively and negatively charged particles dispersed in a permanent magnet, following a design originally made for experiments performed on the Nova-PW laser[60]. The magnet spectrometer is set to analyze particles leaving the plasma along the current sheet (the axis z in our setup), i.e. in the expected direction of the outflow. The diagnostic relies on the deflection of the particles in a well-known magnetic field. The detectors used in this experiment are imaging plates[61].

The HECKLE code
The HECKLE code[63] is a parallel hybrid-PIC code treating ions as macro-particles and electrons as a massless fluid. Macro particles are advanced in time with a leap-frog method and the Lorentz integrator proposed by Boris[64] . Electrons are treated as a massless fluid: their density equals the ions density time the ionization degree at each grid point by quasi-neutrality, their velocity is such as the total current equals the curl of the magnetic field and their pressure evolve in an isothermal way. We hence neglect the transverse component of the displacement current as the phase velocity of electromagnetic fluctuation are small compared to the speed of light. The electric field is provided by the electron's Ohm law, keeping the ideal term, the Hall term (including the total current and the gradient of the electron pressure) and a hyperviscous term to break the magnetic field lines at the scale of the grid. The magnetic fields result from the time integration of the Faraday equation. Electric and magnetic fields are hence calculated self-consistently using a predictor-corrector scheme[65]. The simulation box is periodic in both Z and Y directions. A background ion population (with the same mass, charge and temperature as the foreground ions) with a uniform density equals one fifth of the maximum density in each plasma is superposed, in order to prevent from vacuum region where the calculation of the electric field would then diverge. The following normalizations are made in the code: magnetic field and density are normalized to their maximum initial value (namely $B_0$ and $n_0$) and the mass and charge are normalized to the ones of protons. As a consequence, the lengths, times and velocities are normalized to the ion inertial length, the inverse of ion gyrofrequency and the Alfven velocity (calculated using $B_0$ and $n_0$), respectively. Once the simulation is made, the results are denormalized to SI units following the reverse procedure, which yields the plots and maps shown in Fig.4 and 5.

The grid size is 0.2 (ion inertial length) and the time step is 0.001 (inverse of proton gyroperiod), which is small enough to correctly treat the high frequency whistler modes.

The results of two simulations are shown, namely using a ratio between the plasma and magnetic pressure ($\beta = p_{plasma}/p_{magnetic}$) =1, as well as using $\beta$ =20. These two values are chosen as they correspond to the range of conditions present in our plasma (see Supplementary note 1), and we note that these values also span the ones that can be found along the length solar arches[48]. Using $\beta$=1 makes these results comparable with the very large amount of numerical studies of two-dimensional reconnection, as $\beta \sim 1$ is the one encountered both at the dayside Earth magnetopause[47] and in the Earth plasma sheet[66]. The larger $\beta$ =20 results from the maximum of the initial proton density $n_0$, =10$^{22}$ cm$^{-3}$, the maximum initial proton temperature $T_0$ =1500 eV, and the maximum of the initial magnetic field $B_0$ =600 T retrieved from FCI2 hydro-radiative simulations (see Supplementary note 1). One should also note that $\beta$ *is* calculated using the kinetic pressure of the plasma because the ram pressure is very low, the expanding plasma being drastically slowed-down during its expansion by the inertia of the background (necessary for the hybrid simulations) population.

The ILZ code

This is a test-particle code using a 3D given distribution of electric and magnetic fields, in our case a static map obtained from a snapshot of the HECKLE simulation, to simulate the trajectories of the protons as they pass through and as they are ballistically propagated afterwards, up to the detector. The electromagnetic field is interpolated to the 1$^{st}$ order. The displacement of the particles is governed by the transport equations solved by the Boris[64] algorithm. The protons are considered independent, i.e. they do not interact with each other via Coulomb forces and they do not cause the electromagnetic field to fluctuate. The proton beam is mono-energetic, meaning that we produce synthetic images corresponding to one particular film in the radiochromic film stack. The initial proton distribution is uniform in space and originates from a source point (the solid target T1). When crossing the solid substrate of the targets T2 and T3, the probing protons are scattered, as well as slowed down, in the material. The energy loss is here neglected as the targets are thin and the proton energetic. With respect to scattering, to take its effect into account, we convolve the proton dose calculated by ILZ at the end point on the detector with a Gaussian function that mocks up the scattering according to the Highland formula[67]

$$\theta_{1/e} = \frac{E_s}{p\beta c}\sqrt{\frac{L}{L_R}}\epsilon$$

where $L_R$ is the radiation length in the considered material (here Cu, $L_R$= 1.47 cm), L is the target thickness (here 5 μm), $\beta c$ is equal to the proton velocity ($\beta$ being the standard Lorentz factor), p is the incident proton momentum, $E_s$ = 13.6 MeV is a constant, and $\epsilon$ = 1 + 0.038 log(L/$L_R$) is a corrective factor[68].

**Acknowledgments**
We thank the LULI teams for technical support, and Fabio Reale (INAF) for discussions. This work was supported by the European Research Council (ERC) under the European Unions Horizon 2020 research and innovation program (Grant Agreement No. 787539) and by the Ministry of Education and Science of the Russian Federation under Contract No. 14.Z50.31.0007. This work was partly done within the LABEX Plas@Par project and supported by Grant No. 11-IDEX- 0004-02 from ANR (France). JIHT RAS and NRNU MEPhI members acknowledge the support of RFBR foundation in the frame of projects #14-29- 06099 and #15-32-21121 and the Competitiveness Program of NRNU MEPhI. All data needed to evaluate the conclusions in the paper are present in the paper. Experimental data and simulations are respectively archived on servers at LULI and LPP laboratories and can be consulted upon request. The research leading to these results is supported by Extreme Light Infrastructure Nuclear Physics (ELI-NP) Phase I, a project co-financed by the Romanian Government and European Union through the European Regional Development Fund.

**References**
1. Aulanier, G. *et al.* Slipping Magnetic Reconnection in Coronal Loops. *Science.* **318**, 1588–1591 (2007).
2. Lui, A. T. Y. A synthesis of magnetospheric substorm models. *J. Geophys. Res. Sp. Phys.* **96**, 1849–1856 (1991).
3. de Gouveia Dal Pino, E. M., Piovezan, P. P. & Kadowaki, L. H. S. The role of magnetic reconnection on jet/accretion disk systems. *Astron. Astrophys.* **518**, A5 (2010).
4. Sweet, P. A. The Neutral Point Theory of Solar Flares. in *Electromagnetic Phenomena in Cosmical Physics* **6**, 123 (1958).
5. Parker, E. N. Sweet's mechanism for merging magnetic fields in conducting fluids. *J. Geophys. Res.* **62**, 509 (1957).
6. Ji, H. *et al.* Magnetic reconnection with Sweet-Parker characteristics in two-dimensional laboratory plasmas. *Phys. Plasmas* **6**, 1743–1750 (1999).
7. Loureiro, N. F., Samtaney, R., Schekochihin, A. A. & Uzdensky, D. A. Magnetic reconnection and stochastic plasmoid chains in high-Lundquist-number plasmas. *Phys. Plasmas* **19**, 042303 (2012).
8. Pucci, F. & Velli, M. Reconnection of quasi-singular current sheets: The 'ideal' tearing mode. *Astrophys. J. Lett.* **780**, 4–7 (2014).
9. Hesse, M., Neukirch, T., Schindler, K., Kuznetsova, M. & Zenitani, S. The Diffusion Region in Collisionless Magnetic Reconnection. *Space Sci. Rev.* **160**, 3–23 (2011).
10. Ishizawa, A. & Horiuchi, R. Suppression of Hall-Term Effects by Gyroviscous Cancellation in Steady Collisionless Magnetic Reconnection. **045003**, 2–5 (2005).
11. Fox, W. *et al.* Experimental Verification of the Role of Electron Pressure in Fast Magnetic Reconnection with a Guide Field. *Phys. Rev. Lett.* **118**, 125002 (2017).
12. Katz, N. *et al.* Laboratory Observation of Localized Onset of Magnetic Reconnection. *Phys. Rev. Lett.* **104**, 255004 (2010).
13. Li, C. K. *et al.* Observation of Megagauss-Field Topology Changes due to Magnetic Reconnection in Laser-Produced Plasmas. *Phys. Rev. Lett.* **99**, 055001 (2007).
14. Hare, J. D. *et al.* Anomalous Heating and Plasmoid Formation in a Driven Magnetic Reconnection Experiment. *Phys. Rev. Lett.* **118**, 085001 (2017).
15. Ricci, P., Brackbill, J. U., Daughton, W. & Lapenta, G. Collisionless magnetic reconnection in the presence of a guide field. *Phys. Plasmas* **11**, 4102–4114 (2004).
16. Huang, Y.-M. & Bhattacharjee, A. Turbulent magnetohydrodynamic reconnection mediated by the plasmoid instability. *Astrophys. J.* **818**, 20 (2016).
17. Trattner, K. J., Mulcock, J. S., Petrinec, S. M. & Fuselier, S. A. Location of the reconnection line at the magnetopause during southward IMF conditions. *Geophys. Res. Lett.* **34**, L03108 (2007).
18. Qiu, J., Liu, W. J., Hill, N. & Kazachenko, M. Reconnection and energetics in two-ribbon

flares: A revisit of the bastille-day flare. *Astrophys. J.* **725**, 319–330 (2010).
19. Stanier, A., Simakov, A. N., Chacón, L. & Daughton, W. Fast magnetic reconnection with large guide fields. *Phys. Plasmas* **22**, 010701 (2015).
20. Pritchett, P. L. & Coroniti, F. V. Three-dimensional collisionless magnetic reconnection in the presence of a guide field. *J. Geophys. Res.* **109**, A01220 (2004).
21. Huba, J. D. Hall magnetic reconnection: Guide field dependence. *Phys. Plasmas* **12**, 012322 (2005).
22. Birn, J. & Hesse, M. Energy release and transfer in guide field reconnection. *Phys. Plasmas* **17**, 012109 (2010).
23. Nilson, P. M. *et al.* Magnetic Reconnection and Plasma Dynamics in Two-Beam Laser-Solid Interactions. *Phys. Rev. Lett.* **97**, 255001 (2006).
24. Willingale, L. *et al.* Proton probe measurement of fast advection of magnetic fields by hot electrons. *Plasma Phys. Control. Fusion* **53**, 124026 (2011).
25. Rosenberg, M. J. *et al.* Characterization of single and colliding laser-produced plasma bubbles using Thomson scattering and proton radiography. *Phys. Rev. E* **86**, 056407 (2012).
26. Joglekar, A. S., Thomas, A. G. R., Fox, W. & Bhattacharjee, A. Magnetic Reconnection in Plasma under Inertial Confinement Fusion Conditions Driven by Heat Flux Effects in Ohm's Law. *Phys. Rev. Lett.* **112**, 105004 (2014).
27. Fox, W., Bhattacharjee, A. & Germaschewski, K. Magnetic reconnection in high-energy-density laser-produced plasmas. *Phys. Plasmas* **19**, 056309 (2012).
28. Lancia, L. *et al.* Topology of Megagauss Magnetic Fields and of Heat-Carrying Electrons Produced in a High-Power Laser-Solid Interaction. *Phys. Rev. Lett.* **113**, 235001 (2014).
29. Stamper, J. A. *et al.* Spontaneous Magnetic Fields in Laser-Produced Plasmas. *Phys. Rev. Lett.* **26**, 1012–1015 (1971).
30. Gao, L. *et al.* Precision Mapping of Laser-Driven Magnetic Fields and Their Evolution in High-Energy-Density Plasmas. *Phys. Rev. Lett.* **114**, 215003 (2015).
31. Frank, A. G., Bogdanov, S. Y., Dreiden, G. V., Markov, V. S. & Ostrovskaya, G. V. Structure of the current sheet plasma in the magnetic field with an X line as evidence of the two-fluid plasma properties. *Phys. Lett. A* **348**, 318–325 (2006).
32. Tharp, T. D. *et al.* Quantitative study of guide-field effects on hall reconnection in a laboratory plasma. *Phys. Rev. Lett.* **109**, 2–6 (2012).
33. Zhong, J. *et al.* Modelling loop-top X-ray source and reconnection outflows in solar flares with intense lasers. *Nat. Phys.* **6**, 984–987 (2010).
34. Zhong, J. Y. *et al.* Relativistic Electrons Produced By Reconnecting Electric Fields in a Laser-Driven Bench-Top Solar Flare. *Astrophys. J. Suppl. Ser.* **225**, 30 (2016).
35. Hoshino, M. Stochastic Particle Acceleration in Multiple Magnetic Islands during Reconnection. *Phys. Rev. Lett.* **108**, 135003 (2012).
36. Egedal, J., Daughton, W. & Le, A. Large-scale electron acceleration by parallel electric fields during magnetic reconnection. *Nat. Phys.* **8**, 321–324 (2012).
37. Raymond, A. E. *et al.* Relativistic-electron-driven magnetic reconnection in the laboratory. *Phys. Rev. E* **98**, 043207 (2018).
38. Drake, J. F., Swisdak, M., Che, H. & Shay, M. A. Electron acceleration from contracting magnetic islands during reconnection. **443**, 553–556 (2006).
39. Fox, W. *et al.* Astrophysical particle acceleration mechanisms in colliding magnetized laser-produced plasmas. *Phys. Plasmas* **24**, 092901 (2017).
40. Totorica, S. R., Abel, T. & Fiuza, F. Particle acceleration in laser-driven magnetic

reconnection. *Phys. Plasmas* **24**, 041408 (2017).
41. Wang, S. *et al.* Particle-in-cell simulations of electron energization in laser-driven magnetic reconnection. *New J. Phys.* **18**, 013051 (2016).
42. Rosenberg, M. J. *et al.* A laboratory study of asymmetric magnetic reconnection in strongly driven plasmas. *Nat. Commun.* **6**, 6190 (2015).
43. Lezhnin, K. V. *et al.* Regimes of magnetic reconnection in colliding laser-produced magnetized plasma bubbles. *Phys. Plasmas* **25**, 093105 (2018).
44. Karimabadi, H., Huba, J. D. & Omidi, N. On the generation and structure of the quadrupole magnetic field in the reconnection process : Comparative simulation study. **31**, 3–6 (2004).
45. Smets, R., Aunai, N., Belmont, G., Boniface, C. & Fuchs, J. On the relationship between quadrupolar magnetic field and collisionless reconnection. *Phys. Plasmas* **21**, 062111 (2014).
46. Scurry, L., Russell, C. T. & Gosling, J. T. Geomagnetic activity and the beta dependence of the dayside reconnection rate. **99**, 1–4 (1994).
47. Phan, T. D. *et al.* The dependence of magnetic reconnection on plasma $\beta$ and magnetic shear: Evidence from magnetopause observations. *Geophys. Res. Lett.* **40**, 11–16 (2013).
48. Reale, F. *et al.* 3D MHD modeling of twisted coronal loops. *Astrophys. J.* **830**, 21 (2016).
49. Albertazzi, B. *et al.* Production of large volume, strongly magnetized laser-produced plasmas by use of pulsed external magnetic fields. *Rev. Sci. Instrum.* **84**, 043505 (2013).
50. Cecchetti, C. A. *et al.* Magnetic field measurements in laser-produced plasmas via proton deflectometry. *Phys. Plasmas* **16**, 043102 (2009).
51. Bolton, P. R. *et al.* Instrumentation for diagnostics and control of laser-accelerated proton (ion) beams. *Phys. Medica* **30**, 255–270 (2014).
52. Wilks, S. C. *et al.* Energetic proton generation in ultra-intense laser–solid interactions. *Phys. Plasmas* **8**, 542–549 (2001).
53. Cowan, T. E. *et al.* Ultralow Emittance, Multi-MeV Proton Beams from a Laser Virtual-Cathode Plasma Accelerator. *Phys. Rev. Lett.* **92**, 204801 (2004).
54. Mancic, A. *et al.* Isochoric heating of solids by laser-accelerated protons: Experimental characterization and self-consistent hydrodynamic modeling. *High Energy Density Phys.* **6**, 21–28 (2010).
55. Dromey, B. *et al.* Picosecond metrology of laser-driven proton bursts. *Nat. Commun.* **7**, 10642 (2016).
56. Chen, S. N. *et al.* Absolute dosimetric characterization of Gafchromic EBT3 and HDv2 films using commercial flat-bed scanners and evaluation of the scanner response function variability. *Rev. Sci. Instrum.* **87**, 073301 (2016).
57. Faenov, A. Y. *et al.* High-performance x-ray spectroscopic devices for plasma microsources investigations. *Phys. Scr.* **50**, 333–338 (1994).
58. Revet, G. *et al.* Laboratory unraveling of matter accretion in young stars. *Sci. Adv.* **3**, e1700982 (2017).
59. Rybicki, G. B. & Lightman, A. P. *Radiative Processes in Astrophysics*. *Physics Bulletin* **31**, (Wiley-VCH Verlag GmbH & Co. KGaA, 1985).
60. Cowan, T. E. *et al.* High energy electrons, nuclear phenomena and heating in petawatt laser-solid experiments. *Laser Part. Beams* **17**, 773–783 (1999).

61. Boutoux, G. *et al.* Study of imaging plate detector sensitivity to 5-18 MeV electrons. *Rev. Sci. Instrum.* **86**, 113304 (2015).
62. Lelasseux, V. & Fuchs, J. *No Title*.
63. Smets, R., Belmont, G., Aunai, N. & Rezeau, L. Plasma diffusion in self-consistent fluctuations. *Phys. Plasmas* **18**, (2011).
64. Boris, J. P. Relativistic plasma simulation—Optimization of a hybridcode. in *Proceedings of 4th Conference on Numerical Simulation of Plasmas* 3–67 (1970).
65. Harned, D. S. Quasineutral hybrid simulation of macroscopic plasma phenomena. *J. Comput. Phys.* **47**, 452–462 (1982).
66. Lui, A. T. Y. A synthesis of magnetospheric substorm models. *J. Geophys. Res. Sp. Phys.* **96**, 1849–1856 (1991).
67. Highland, V. L. Some practical remarks on multiple scattering. *Nucl. Instruments Methods* **129**, 497–499 (1975).
68. Groom, D. E. & Klein, S. R. Passage of particles through matter. *Eur. Phys. J. C* **15**, 163–173 (2000).